\def\year{2022}\relax
\documentclass[letterpaper]{article} 
\usepackage{aaai23}  
\usepackage{times}  
\usepackage{helvet}  
\usepackage{courier}  
\usepackage[hyphens]{url}  
\usepackage{graphicx} 
\usepackage{natbib}  
\usepackage{caption} 
\usepackage{algorithm}
\usepackage{algorithmic}

\usepackage{newfloat}
\usepackage{listings}
\DeclareCaptionStyle{ruled}{labelfont=normalfont,labelsep=colon,strut=off} 
\floatstyle{ruled}
\newfloat{listing}{tb}{lst}{}
\floatname{listing}{Listing}
\usepackage{subcaption}
\usepackage{tabularx}
\usepackage{booktabs}
\usepackage{comment}
\usepackage{multirow}
\usepackage{amsmath}
\usepackage{soul}
\usepackage{xspace}
\usepackage{xcolor,colortbl}

\definecolor{Gray}{gray}{0.85}
\definecolor{LightCyan}{rgb}{0.88,1,1}

\newcolumntype{a}{>{\columncolor{Gray}}r}
\usepackage{xcolor}

\title{Happenstance: Utilizing Semantic Search to Track Russian State Media Narratives about the Russo-Ukrainian War On Reddit}
\author {
    Hans W. A. Hanley, 
    Deepak Kumar, 
    Zakir Durumeric\\
}
\affiliations {
    Stanford University \\
    hhanley@stanford.edu, kumarde@stanford.edu, zakird@stanford.edu
}

\begin{document}

\maketitle

\begin{abstract}
In the buildup to and in the weeks following the Russian Federation's invasion of Ukraine, Russian state media outlets output torrents of misleading and outright false information. In this work, we study this coordinated information campaign in order to understand the most prominent state media narratives touted by the Russian government to English-speaking audiences. To do this, we first perform sentence-level topic analysis using the large-language model MPNet on articles published by ten different pro-Russian propaganda websites including the new Russian ``fact-checking'' website waronfakes.com. Within this ecosystem, we show that smaller websites like katehon.com were highly effective at publishing topics that were later echoed by other Russian sites. After analyzing this set of Russian information narratives, we then analyze their correspondence with narratives and topics of discussion on \texttt{r/Russia} and 10 other political subreddits. Using MPNet and a semantic search algorithm, we map these subreddits' comments to the set of topics extracted from our set of Russian websites, finding that 39.6\% of \texttt{r/Russia} comments corresponded to narratives from pro-Russian propaganda websites compared to 8.86\% on \texttt{r/politics}.
\end{abstract}

\vspace{-10pt}

\section{Introduction}

On February 24, 2022, the Russian Federation invaded Ukraine. As reported by NBC News, in the weeks leading up to the war and in the days following the invasion, Russian disinformation campaigns targeting Ukraine and blaming the ``West'' for heightening tensions increased dramatically~\cite{Abbruzzese2022}. Narratives ranged from the debunked idea that the US was funding biological weapons research in Ukraine to the claim that Russia waged the war to ``demilitarize and denazify'' Ukraine. As a result of this torrent of disinformation, the United Kingdom and the European Union banned Russian media companies like Russia Today~\cite{M2022, Chee2022}. On March 1, due to the vast amounts of disinformation, Reddit even took the extraordinary step of quarantining the \texttt{r/Russia} subreddit~\cite{Yeo2022}. This quarantine amounted to publicly labeling \texttt{r/Russia} as containing disinformation and requiring users to acknowledge this before accessing the subreddit.


However, despite the prevalence of this type of disinformation online, the research community still lacks programmatic approaches for tracking the spread of specific disinformation narratives---like those about Ukraine---across both news sites and social media platforms. Topic modeling tools like LDA fall short in mapping topics across platforms~\cite{min2015cross}, and keyword-based approaches often rely on pre-existing expert knowledge of disinformation campaigns, which often cannot be distilled at the speed at which information campaigns are deployed~\cite{bal2020analysing}.


To address these limitations, in this paper, we validate and utilize a \textit{sentence-level} topic analysis methodology to identify and map the spread of Russian state media narratives across news sites and social media. Specifically, our approach leverages the large-language model MPNet's understanding of English to embed sentences to a high-dimensional subspace~\cite{song2020mpnet}. Once mapped, as in BERTopic~\cite{grootendorst2020bertopic}, we utilize the dimensionality reduction algorithm UMAP~\cite{becht2019dimensionality}, the density-based clustering algorithm HDBSCAN~\cite{mcinnes2017hdbscan}, and finally class-based TF-IDF~\cite{ozgur2005text} to extract topic keywords. We note that this methodology, based on works like BERTopic~\cite{grootendorst2020bertopic} and Top2Vec~\cite{angelov2020top2vec} and vital to how we later understand the spread of the identified topics across social media, is based on the assumption that each item in the analyzed dataset is about \textit{one topic}. We, therefore, utilize the intuition that each sentence in a news article is about only \textit{one topic} and take a \textit{sentence-level} topic analysis throughout this work. Using this methodology, we analyze the \textit{topics/narratives} promoted by ten Russian state media websites~\cite{RussiaPillar2020} including the new ``fact-checking'' website waronfakes.com between January 1 and April 5, 2022. 


We show that several disinformation narratives were widely reported and referenced in dozens of articles across each of our scraped Russian websites. For instance, we document that roughly 35~Russia Today and 44~Sputnik News articles pushed the debunked theory that the US-funded biological weapons laboratories in Ukraine~\cite{Price2022}. We further observe that several key websites are responsible for introducing and propagating state media narratives. For instance, whenever the website katehon.com introduced a new narrative, other Russian websites in our dataset produced an average of 21~additional articles about the same topic.  
After understanding the narratives promoted by our set of pro-Russian propaganda websites, we then study these narratives' influence on the \texttt{r/Russia} subreddit. Using MPnet, we map \texttt{r/Russia} comments to the same dimensional subspace as the sentences from our set of Russian state media websites. Using the assumption that each news article sentence and each Reddit comment is about \textit{one topic}, utilizing  MPNet we identify the cluster of news article sentences that have the highest semantic similarity to each Reddit comment; essentially performing \textit{semantic search}. Thresholding to ensure that each comment has a high minimum semantic similarity to its matched cluster of news article sentences and using the cluster's TF-IDF topic labels, we thus match Reddit comments to previously identified Russian state media narratives. This approach enables us to identify Reddit comments that are about the same topics as those propagated by Russian media outlets without having to depend on keywords \emph{and} while taking into account synonyms and semantic variants of the words within our pre-identified topic clusters.

With this approach, we find that 39.6\% of the comments on \texttt{r/Russia} between January~1 and March~15, 2022, discussed topics/narratives published by Russian state media sites. Mapping an additional 5.37~million comments from 10~other political subreddits to the same embedding space and calculating their percentage of comments associated with Russian state media, we programmatically show that \texttt{r/Russia} had elevated levels of Russian state media-associated comments compared to other subreddits (\textit{i.e} 8.86\% on \texttt{r/politics}). Showcasing our approach's ability to track disinformation, we finally track the spread of two specific Russian disinformation narratives across all 11~documented subreddits. 

Our case study shows that sentence-level language analysis is an effective methodology for programmatically identifying and tracking news narratives as they spread across platforms. We hope that it can serve as the basis for future studies about online disinformation.



\vspace{-5pt}

\section{Related Work}

\paragraph{Russian Disinformation}
The Russian government has conducted information warfare throughout its history~\cite{jowett2018propaganda}. In the past decade, however, the amount of disinformation spread by the Russian Federation has increased substantially~\cite{hellman2017can}. Due to this increase, as well as Russian interference in the 2016 US Presidential elections~\cite{badawy2019characterizing}, there have been multiple studies of Russian-spread disinformation on social media platforms. For example, Badawy et~al.\ studied the effect of Russian government-linked misinformation bots and trolls on Twitter~\cite{badawy2018analyzing}. Similarly, Golovchenko et~al.\ found that a large majority of the message on Twitter promoting pro-Russian narratives surrounding events in Ukraine belonged to ostensibly non-state-sponsored accounts~\cite{golovchenko2018state}.

In addition to studies on the spread of Russian disinformation on social media, several other works have documented the general spread of news misinformation on social media. Most similar to our work, Guo et~al.\ attempted to link tweets to news articles using text-to-text correlations~\cite{guo2013linking}. While unable to handle synonyms and perform larger topic correlations between articles, their approach largely informs our own. In a similar vein, Liu et~al.\ mine Weibo and Twitter to identify the spread of misinformation around events like the downing of flight MH\,370 in Ukraine~\cite{liu2018mining}.

\paragraph{Topic Modeling}
Our work largely depends on identifying and performing topic analysis for comments and sentence-level texts and there has been significant prior work on topic modeling for short texts. Latent Dirichlet Allocation (LDA), a Bayesian probabilistic model used to assign topics to documents, is one of the most commonly used methodologies for extracting topics~\cite{jelodar2019latent}. In their work, Albalawi et~al.\ show that LDA is one of the most effective methodologies among various computationally light alternatives (e.g., LSA, LDA, NMF, PCA, RP) proposed within the last decade based on metrics of recall and precision of topics for short text data~\cite{albalawi2020using}. However, due to the problem of sparsity in word co-occurrences, LDA often falls short. Qiang et~al.\ discuss the positives and negatives of many topic modeling approaches, highlighting LDA's shortcomings~\cite{qiang2020short}. 

Several recent works have shown the usefulness of word embeddings in improving LDA-based approaches. Finding LDA unable to deal with large vocabularies, Dieng et~al.\ extend LDA by building topics directly from word-embedding spaces~\cite{dieng2020topic}. As in our work, Top2Vec~\cite{angelov2020top2vec} and BERTopic~\cite{grootendorst2020bertopic} both utilize word embeddings, followed by UMAP and HDBScan identify topics. We note that our use of BERTopic's approach largely falls into the word embedding topic clustering approaches utilized by large social media companies to group together similar articles~\cite{qiang2020short}. Finally, the MPNet authors found that they could achieve better results on similarity and semantic search tasks on single sentences (key aspects when performing topic clustering) utilizing a model that accounts for auxiliary position information. This allows the MPNet to consider the full sentence being transformed during training~\cite{song2020mpnet}. Their work has enabled increased MPNet's use for semantic search and topic analysis~\cite{huertas2021countering} and we use MPNet in our work.

\paragraph{Tracking Online News and Media}
As the influence of online media has grown, several works have examined the spread of ideas, memes, and topics within and across platforms. Leskovec et~al.\  utilize a clustering approach based on directed acyclic graphs to identify and trace the growth of particular ``memes'' across over 1.65 million blogs and news sites~\cite{leskovec2009meme}. Gomez-Rodriguez et~al.\ adopt a cascade transmission model to identify information diffusion patterns and influence of particular websites by utilizing data from 170 million blogs and articles~\cite{gomez2012inferring}. Myers et~al.\ utilize similar approaches to understand the different types of diffusion and topic propagations across social media websites~\cite{myers2012information}. In a similar vein, Zannettou et~al.\ track the memes from fringe online communities~\cite{zannettou2018origins}. Finally, utilizing hyperlink, image, and qualitative analysis techniques, Starbird et~al.\ examine the spread of rumors during crisis events and Hanley et~al.\ document the spread of the QAnon conspiracy theory~\cite{starbird2018ecosystem,hanley2021no}.


\vspace{-10pt}
\section{Methodology}
\label{sec:methodology}

To conduct our analysis of Russian state media websites and their impact on social media conversations, we collected two datasets: (1) article texts published by Russian state-sponsored media websites, and (2) submissions and comments from \texttt{r/Russia} and 10~other political subreddits~\cite{rajadesingan2020quick}. In this section, we detail each of these datasets and discuss our methodology for performing topic analysis.

We note that throughout this work we refer to topics/narratives extracted from our set of Russian state media and propaganda websites as ``Russian state media topic/narratives'' while referring to specific debunked stories as ``disinformation'' in line with prior work. We do this because while not all stories from these Russian propaganda websites are necessarily false, all are still state-promoted. Conversely, specific narratives that are known to be false but are promoted by these propaganda sites for specific political purposes, we consider falling under the narrower definition of ``disinformation''~\cite{jack2017lexicon}.
\begin{figure}
  \centering
  \includegraphics[width=1\linewidth]{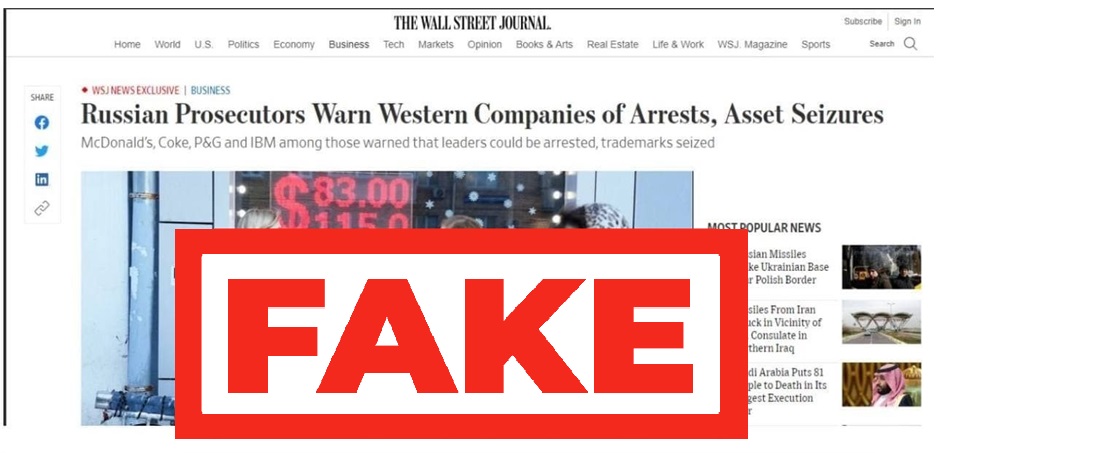}
\caption{WarOnFakes.com ``fact-checks'' the Wall Street Journal reports that the Russian government plans to seize the assets of companies that leave the Russian market. }
\label{fig:waronfakes-examples}
\vspace{-5pt}
\end{figure}
\vspace{-5pt}
\subsection{Articles from Russian State Media}
Our study examines nine English-language Russian propaganda and state media websites documented by the US State Department~\cite{RussiaPillar2020}: rt.com (RT, 470~articles), sputniknews.com (SN, 519), strategic-culture.org (SC, 85), journal-neo.org (JN, 79), news-front.info (NF, 361), katehon.com (KH, 62),  geopolitca.ru (GP, 73), southfront.org (SF, 193), and tass.com (Tass, 674). We further include the recently launched site waronfakes.com (WoF, 167). Purportedly run by journalists and experts, waronfakes.com began publishing articles ``fact-checking'' news and statements from Western media as well as NATO-aligned politicians (Figure~\ref{fig:waronfakes-examples}). The website has been promoted by the Russian Embassy in the US.\footnote{\url{https://web.archive.org/web/20220514013851/https://twitter.com/mfa_russia/status/1500223302941487107}} The New York Times further investigated the site and found it to be a hub of Russian disinformation about the war in Ukraine~\cite{Thompson2022}. The website has published several articles denying Russian war crimes in the Kyiv suburb of  Bucha\footnote{\url{https://web.archive.org/web/20220408210312/https://waronfakes.com/lies-about-bucha/fake-bodies-of-civilians-have-been-lying-on-the-streets-of-bucha-since-march-11/}} and the city of Kramatorsk.\footnote{\url{https://web.archive.org/web/20220408210347/https://waronfakes.com/civil/russian-army-hit-the-railway-station-in-kramatorsk-with-a-missile/}}

For each website, we collect the set of articles the website published in 2022 about Ukraine. To do this, we crawl each website using {Selenium}. After visiting each site's homepage, we use a breadth-first approach to find articles that mention Ukraine in their article body. We scrape 5~hops from the root page (i.e., we collect all URLs linked from the homepage [1st hop], then all URLs linked from those pages [2nd hop], and so forth). We further supplement this corpus by using Google's API to find and add articles indexed in 2022 that mention Ukraine for each site. To extract article information from each page, we use the Python \texttt{newspaper3k} library and extract article publication date using the \texttt{htmldate} library~\cite{barbaresi2020htmldate}. In total, we collected 2,683~unique articles.

\vspace{-5pt}
\subsection{Reddit Dataset} To understand the spread of Russian state media narratives about the Russo-Ukrainian war on Reddit, we collect posts and comments posted on \texttt{r/Russia} from January 1, 2022, to March 15, 2022, using Pushshift~\cite{baumgartner2020pushshift}, which keeps a queryable replica of Reddit data. However, due to a Pushshift outage, we directly collected submissions and comments from March 1--15 via Reddit's API\@. Altogether, this dataset consists of 101,122~Reddit comments and 6,984~Reddit submissions from \texttt{r/Russia}. To later validate our approach, we collect an additional 5.37~million comments posted between January 1 and March 15 from 10~other political subreddits~\cite{rajadesingan2020quick} including \texttt{r/politics} and \texttt{r/conservative} (Table~\ref{tab:all-top-narratives-reddit}).

\begin{figure}
  \centering
  \includegraphics[width=1\linewidth]{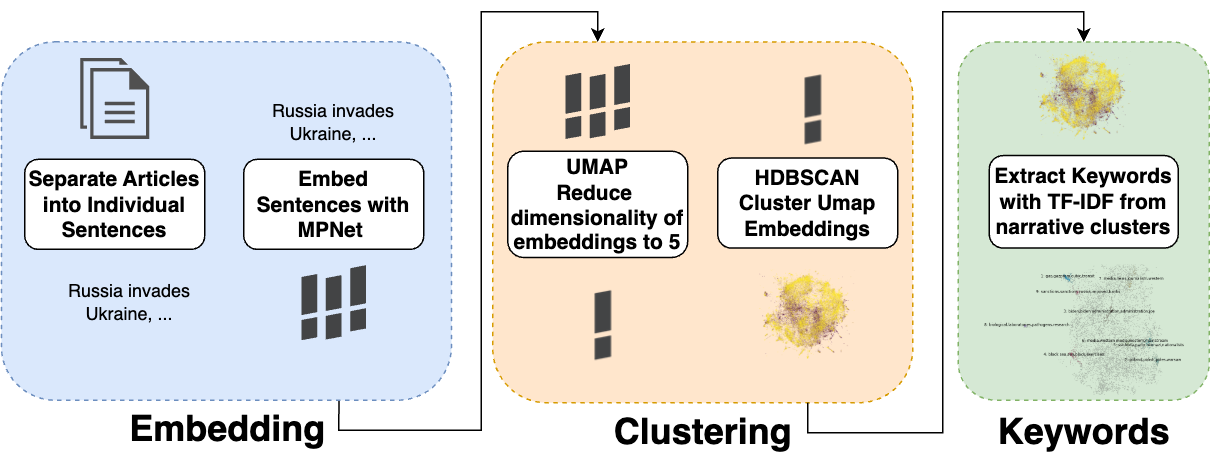}
  \caption{
  We extract topics from online articles by dividing the text into individual sentences, embedding them in a 768-dimensional subspace, reducing their dimensionality using UMAP, clustering them using HDBScan, and finally extracting keywords using class-based TF-IDF\@. The clustering and keyword extraction follows the methodology specified in BERTopic~\cite{grootendorst2020bertopic}.
  }
\label{fig:topic-modeling}
\vspace{-0.5cm}
\end{figure}

\subsection{Topic Extraction and Analysis} 

To extract topics and higher-level semantic meaning from our documents and comments, we rely on recent advancements in contextual word embeddings~\cite{huang-etal-2021-whiteningbert-easy,devlin2018bert,song2020mpnet}. Notably, these advancements allow text with similar meanings, when mapped to a given embedding space, to have similar embeddings. Our work leverages these advancements to build a semantically-rich embedding space for sentences from Russian state media websites, to cluster these sentences into narrative/topic clusters, and to finally extract topics from these clusters. We note that our approach relies on each embedding belonging to only \emph{one cluster}, and thus to \emph{one topic}. We thus rely on \textit{sentence-level} topic analysis in this work. Our intuition is that while articles often contain multiple topics, sentences individually tend to discuss one topic and thus can be more accurately placed into only one topic/narrative cluster. Figure~\ref{fig:topic-modeling} shows each step of our pipeline. We detail each step below:

\paragraph{Extracting Sentence-Level Embeddings} The first step of our topic modeling pipeline leverages a fine-tuned version of MPNet~\cite{song2020mpnet}, which is a 768-dimensional contextual word embedding model released by Microsoft Research. We leverage MPNet's understanding of English to group together sentences (and therefore, articles) that share a high semantic similarity. The specific model version we use is fine-tuned to the \textit{semantic search task}, which aims to find documents that relate to one other~\cite{guha2003semantic}. 

To prepare each article for input to MPNet for topic analysis, we utilize the Python \texttt{Natural Language Toolkit/nltk} library's sentence tokenizer to segment the articles into their individual sentences~\cite{loper2002nltk}. We further remove all special characters, hyperlinks, and non-English words.

\paragraph{Forming Topic/Narrative Clusters} After extracting each sentence-level embedding, we create ``narrative clusters'' by clustering similar sentences. To do this, we first perform dimensionality reduction using the UMAP algorithm~\cite{becht2019dimensionality}, which reduces the dimensionality of our embeddings from 768 to 5. We perform this dimensionality reduction to avoid the ``curse of dimensionality'' that makes it difficult to identify dense clusters or perform nearest neighbor searches in high-dimensional spaces at a reasonable computational cost ~\cite{marimont1979nearest}. We then cluster our set of embedding using hierarchical density-based clustering, namely HDBSCAN~\cite{mcinnes2017hdbscan}. We note that HDBSCAN is useful to our methodology as it allows us to identify clusters of arbitrary size. Furthermore, HDBSCAN allows us to identify topics without us pre-defining the number of clusters, enabling us to find the ``naturally'' occurring dense narrative groupings within our dataset~\cite{zannettou2018origins}. 

HDBSCAN is conservative, assigning sentences and embeddings to a cluster only when confidence is high~\cite{mcinnes2017hdbscan}. As a result after clustering, a significant percentage of the data is categorized as outliers (in our case, 33.7\% of Russian article sentences). These outliers are sentences that are mere ``one-off'' ideas that often do not appear repeatedly. For example, in an article criticizing the U.S. government for its concerns about Russia potentially using chemical weapons in Ukraine, a Russian state media website mentioned that the US had used ``Agent Orange in Vietnam'' and thus the US' concern was hypocritical. This specific sentence was not part of a consistent narrative across our websites and was thus considered an outlier. We utilize the default parameters outlined in Grootendorst et~al. for our clustering and dimensionality reduction. We perform a formal evaluation of this methodology in Section~\ref{sec:topics}.

\paragraph{Extracting Important Keywords} For each of our narrative clusters, we perform keyword extraction with class-based TF-IDF. We extract the top 10~key unigram and bigrams from the sentences in each cluster using TF-IDF\@.

\vspace{-5pt}

\section{Narrative Evolution in Russian Media} 
\label{sec:topics}
In this section, we use our topic analysis technique from Section~\ref{sec:methodology} to cluster sentences from 10~Russian state media websites into topic clusters. We then show how these topic clusters can be used to measure how far topics spread and the influence of specific actors in Russian state media.

\begin{table}
\centering
\fontsize{9pt}{7.5pt}
\selectfont
\setlength{\tabcolsep}{4pt}
\begin{tabular}{llrrr}
{} & Keywords &{\# Artl.}  &{\# Sent. } & {Prec.}           \\  \midrule 
1 & joe biden  &204 & 520 &  100     \\ 
2 & biological,laboratories  &200 & 749 &  97.5      \\
3 & negotiations,talks &200 & 387 &  99.4     \\ 
4 & media, western  &182 & 417 & 100      \\
5 & sanctions, individuals  &131 & 210 &  100      \\
6 & join nato  &116 & 143 &  100      \\ 
7 & kyiv forces  &112 & 278 &  97.2      \\
8 & gas, russia gas  &100 & 419 &  99.3      \\ 
9 & demilitarize, denazification  &96 & 120 & 100     \\ 
10 & evacuate,evacuation  &89 & 101 &  96.0     \\ \midrule 
14 & bloody crimes  &81 & 85 &  97.6      \\ 
50  & emotions, inexcusable & 57 &74 &95.9          \\ 
111  & operation, responses &  48 & 48& 100          \\ 
16 & embassy, evacuate & 74& 140& 98.6   \\  
264 & russian borders  &  27 & 28  &  100     \\
127  & cuba, kennedy &  17 & 44 & 100        \\   
122 & boris, scandals &  18 & 45 & 100           \\  
382 & contact, heavy shelling  &20 &22 &100            \\ 
378 &invades, inadvertent, bloomberg &  15 &   22 & 100        \\ 
26  &serbia, yugoslavia & 43 & 108   & 98.1        \\ 
312  &ukraine lose & 20 & 26   & 96.1        \\  %
 781 &territory heartland & 12 & 13   & 100       \\ %
281  &launched special & 35 & 35   & 100       \\ %
191  &delegation, belarus & 17 & 23   & 100       \\ 
160  &bucha, crimes, withdrew & 27 & 38   & 97.3       \\  \midrule 
    \multicolumn{4}{r}{Overall Precision:}    &{{98.9}} \\ 
\end{tabular}
\caption{\label{tab:precision-test} Evaluation of the precision of our topic analysis model on 25 topics (top 10 most frequently mentioned topics and 15 random topics) derived from Russian state media website articles. }  
\vspace{-0.5cm}

\end{table}
\vspace{-5pt}
\subsection{Evaluating the Topic Model}
From our articles, after extracting embeddings and reducing their dimensionality with UMAP, HDBSCAN identified 1056 different topics with 753 separate topics having more than 10 articles mentions. The median article's sentences belong to five different clusters. However, before discussing the largest topics and the interaction of websites within our dataset, we first evaluate our methodology.

To begin, we compute a topic coherence metric, which is a proxy for how ``human-understandable'' the generated topics are. We compute the word2vec coherence metric and find a coherence of 0.563 (scaled from [0--1] with the top 10~unigrams and bigrams)~\cite{o2015analysis}. Considering the proportion of unique words in each topic~\cite{dieng2020topic} as a metric for topic diversity, we achieve a topic diversity score of 0.874 (again scaled from [0--1] with the top 10~unigrams and bigrams), which establishes that each of our topics on average contains terms that are highly unique to itself.

Next, having seen that our topics are coherent and diverse, we compute the average inter-cluster cosine similarity, which determines how similar the sentences within our clusters are. This ensures that each cluster contains sentences that are about the same topic. We see a score overall average score of 0.560 on a scale [0,1]. For context, the sentence ``\textit{Has humanity really, with fewer and fewer exceptions, fallen into the complete darkness of hedonism, conformism, moral and spiritual blindness}'' and the sentence ``\textit{How is it possible that as a planetary collective, as humanity as a whole, we have not seen for a moment the greatest deception of all time and that by our inaction we agree to be complicit in our own destruction: moral, spiritual, intellectual, and at the end, physical}'' (both from an article published on geopolitica.ru about the need to support Russia in the Russo-Ukrainian war) have a similarity of 0.58. Collectively, our results illustrate that sentences within each cluster are similar and that the cluster topics are coherent and diverse.

\begin{table}
\centering
\fontsize{9pt}{7.5pt}
\selectfont
\setlength{\tabcolsep}{4pt}
\begin{tabular}{lrrr}
Domain & Origin &  Avg.\ Origin & Avg.\ Non-Origin \\ 
& Topics  &  Articles & Articles\\ \midrule
rt.com &\textbf{287} & 4.97 & 3.84\\
news-front.info &216& 4.10& 2.86\\
strategic-culture.org&112 &3.02 &2.74 \\
tass.com &121 & \textbf{7.86}& \textbf{4.14}\\
katehon.com &108 &2.71&  1.86\\
geopolitica.ru &100&3.78 &2.52 \\
southfront.org &97 &5.78 &3.26\\
journal-neo.org&68  &3.21 & 2.17 \\
sputniknews.com&64  &2.56 & 2.78  \\
waronfakes.com &21 &4.71 & 2.32\\
\end{tabular}
\caption{\label{tab:real-originating} Number of originating topics on each domain and the average number of external articles written about domain-originating topics vs non-originating topics. }  
\vspace{-5pt}
\end{table}
Finally, we analyze the accuracy of our clustering by investigating whether the topics assigned to a cluster accurately reflect the news article sentences in the cluster. To do this, in addition to the top 10 most frequently mentioned topics, we take a random sample of 15~topics and determine the fraction of sentences that accurately conform to the extracted topics (Table~\ref{tab:precision-test}). Specifically, one expert manually verified if each sentence within the cluster matched the topic indicated by the TF-IDF keywords. Altogether 4,095 sentences were examined across the 25 different topics. Each cluster that was tested contains sentences that conform to the given TF-IDF extracted topic keywords with a precision of at least 95.9\%. As an example of an error, the sentence \textit{``Omicron is sneaky because it has symptoms of a common cold: runny nose, slight cough, lack of temperature, said Klitschko, explaining that he has now tested negative.''} was classified as being part of the biological laboratories cluster (Topic~2).

We note that this approach enables extracting granular stories/narratives. For example, on February 4, 2022, prior to Russia invading Ukraine, the news website Bloomberg accidentally published a headline saying that Russia had invaded Ukraine~\cite{Bloomberg}. This story (Topic 378) was largely derided by the news websites in our dataset with 15 different articles published about the incident by our set of Russian websites. Similarly, Russian war crimes in the city of Bucha in Ukraine, widely covered in the Western press, were also noted in 27 different articles in our dataset. Looking at these articles, we see our cluster picked up on the debunked disinformation narrative~\cite{Browne2022} that the Russian military had withdrawn from Bucha before the atrocities began. While not large stories in our dataset, our approach was able to cluster and identify both, illustrating its ability to detect small but important narrative threads. 
\begin{figure}
  \centering
  \includegraphics[width=1.0\linewidth]{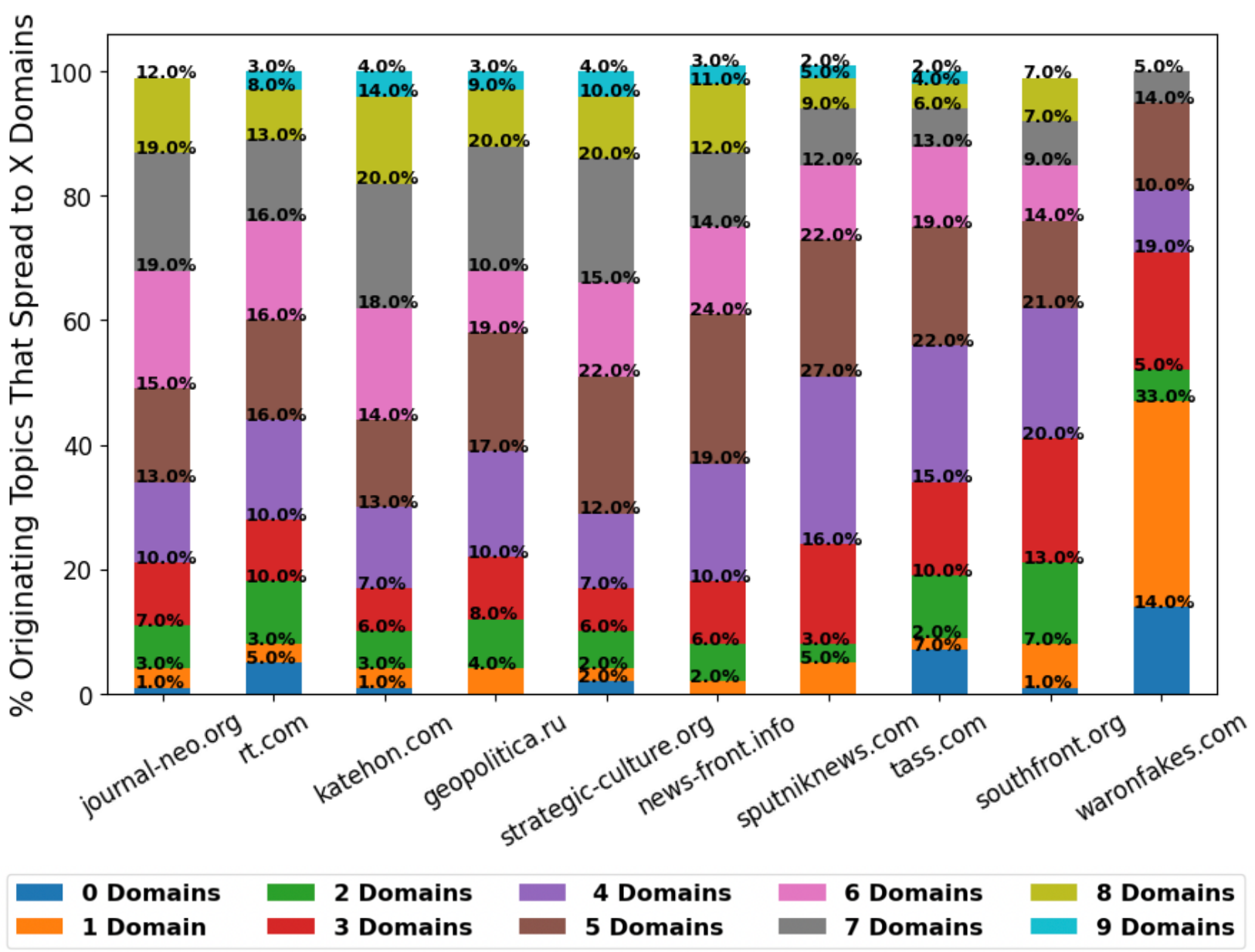}
\caption{Percentage of Originating Topics that spread to X domains. Each domain's originating topics spread to many of the other Russian state media websites.
}
\label{fig:narrative-spread}
\vspace{-5pt}
\end{figure}
\vspace{-5pt}
\subsection{Origins of Topics in Russian State Media }
Each website in our dataset originates several topics/narratives. We consider a website to originate a narrative if it published an article containing the topic on the first day that the topic appeared in our dataset (more than one website can originate a topic). Table~\ref{tab:real-originating} shows the number of originating articles for each state media website. Rt.com originates the most topics, while waronfakes.com (a newly created website) originates the fewest. Most topics that begin on a site travel widely throughout the Russian state media ecosystem (Figure~\ref{fig:narrative-spread}). For example, 82\% of rt.com topics about Ukraine eventually spread to at least three other sites. In only one case---waronfakes.com---do we see fewer than 50\% of topics propagate to at least three~other websites.

As seen in Table~\ref{tab:originating}, several of the smaller state media websites in our dataset originated topics that were then subsequently published widely within the ecosystem. Katehon.com in particular has a large sway. Whenever the site originated a topic, an average of 21.32~external articles were written about the topic. Similarly, when newsfront.info and geopolitica.ru originated a topic or narrative, other websites wrote an average 17.59 and 18.91~articles on the topic.
\begin{figure}
  \centering
  \includegraphics[width=1\linewidth]{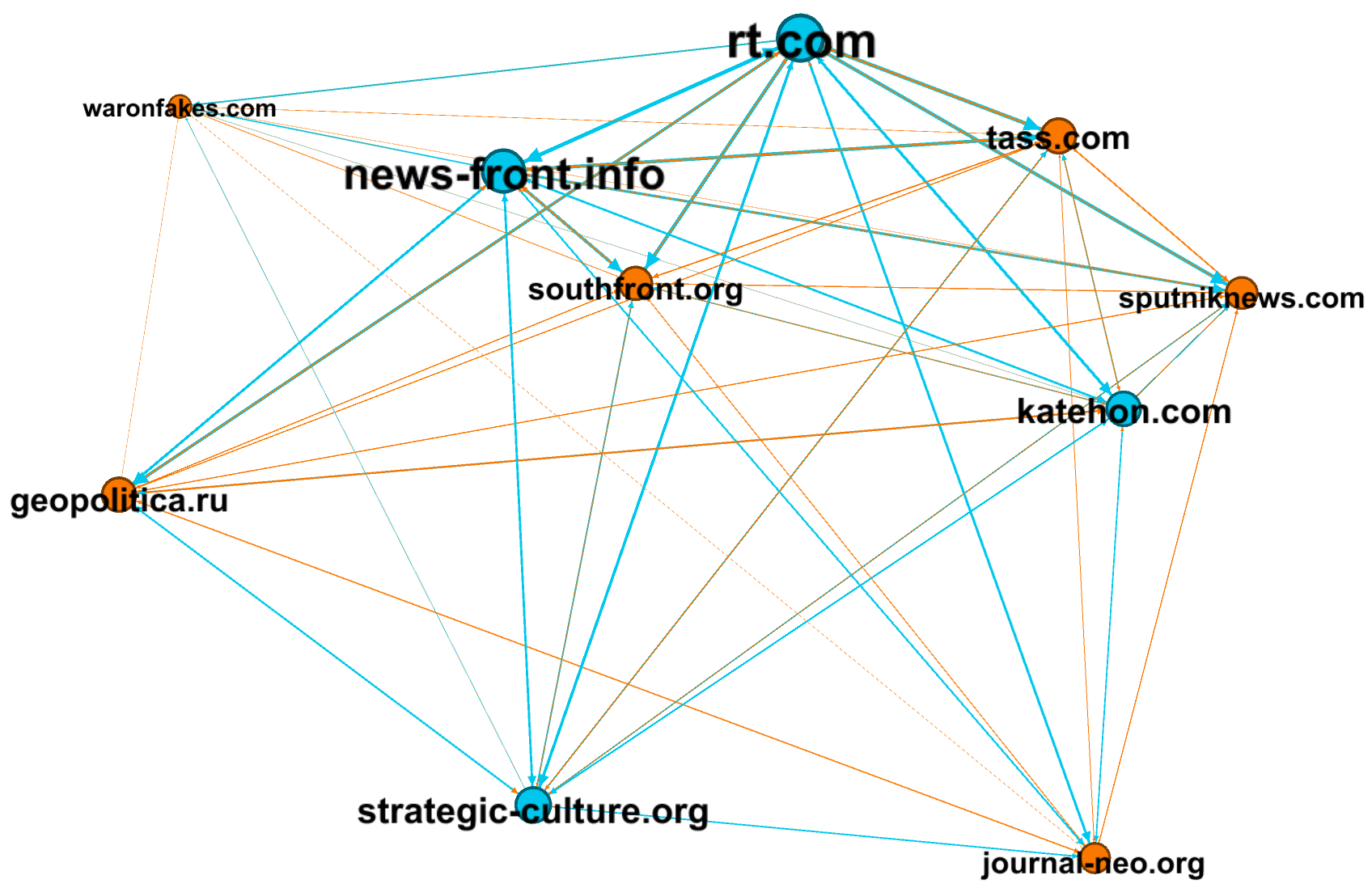}
\caption{Relationships between different Russian state media websites. Node size is determined by weighted out-degree (number of topics/narratives originated) that were echoed by other websites. Websites that broadcast more stories than they echo are in blue; websites that echo more topics/narratives than they originate are in dark orange. 
}
\label{fig:story-sink}
\vspace{-10pt}
\end{figure}

We visualize the interconnections and relationships between our set of Russian websites based on the sharing of topics. This helps us understand where topics that originated on a given site eventually migrate. In Figure~\ref{fig:story-sink}, we draw a weighted directed edge from an origin website to another website based on the amount of originating topics/narratives that were subsequently written about on the receiving website. We color code websites that originate more narratives than they echo as blue; websites that echo more narratives than they originate are color coded as dark orange. As seen in Figure~\ref{fig:story-sink}, rt.com, news-front.info, katehon.com, and strategic-culture.org are some of the most prominent in generating and propagating narratives within this ecosystem. We note that the latter three are also some of the most widely echoed in the ecosystem (Table~\ref{tab:originating}), evidencing these particular sites' ability to lead and control narratives in the Russian information space.

We observe that almost every website in our dataset writes more articles about the topics that they originate. We confirm this with a Mann-Whitney U-test, using a p-value of 0.005 (p-value of 0.05 with a Bonferonni correction of 10), and we find significant results for every website except sputniknews.com. This indicates---with the exception of sputniknews.com, which appears to be more of a receptacle of narratives---that when a website introduces a new narrative/topic, it promotes it more vigorously.

Next, we investigate whether the number of external articles published about a topic correlates with the number of articles published by the originating website. This allows us to more closely examine whether the originating website's promotion of a topic correlates with external websites writing more about that topic. For a website whose correlation's corresponding p-value determined with a t-test is non-significant ($>0.05$), we do not report the correlation. Looking at the correlations, while most of the correlations were insignificant, we do see that as newsfront.info, strategic-culture.org, geopolitica.ru, and kaethon.com publish more on their originating topics, other websites write more articles about these topics. This correlation is strongest for the think-tank website strategic-culture.org with $\rho=0.58$.
\begin{table}
\centering
\setlength\tabcolsep{4pt}
\fontsize{9pt}{7.5pt}
\selectfont
\begin{tabular}{lrr}
Domain &  Avg.\ External  & Corr. \\  
       & Articles Per Topic \\ \midrule
rt.com & 13.94 & --\\
news-front.info &  17.59 & 0.525\\ 
strategic-culture.org &  16.59 & \textbf{0.580}\\ 
tass.com &  10.98 & --\\ 
katehon.com &  \textbf{21.32}  &0.408 \\ 
geopolitica.ru & 18.91& 0.472 \\ 
southfront.org & 10.46 & -- \\ 
journal-neo.org & 13.21 & -- \\ 
sputniknews.com &  12.30 & -- \\ 
waronfakes.com & 4.81 & -- \\ 
\end{tabular}
\caption{\label{tab:originating} Average number of external articles about each website's originating topics and their Pearson correlation with the number of on-site articles written about that topic. We show correlations only when the p-value $\leq 0.05$.}  
\vspace{-10pt}
\end{table}


\begin{table*}
\centering
\fontsize{9.0pt}{8.0pt}
\selectfont
\begin{tabular}{l|l|rrrrrrrrrr}
 &  {Keywords} & RT  & SN & NF  &  Tass &  SC  &  KH  &  GP  &  JN &SF & WoF \\ \midrule
 1 & biden, joe, joe biden, president joe, us president
 & \textbf{44} & 35& 29&23 &20 &8 &13 &19 & 13 & 0  \\
 2 & biological, laboratories, pathogens, research, chemical & 35 &\textbf{44}&19 &37 &13 &5 &11 & 11 &20 & 5\\
 3 & negotiations, talks, delegation, russian delegation, ukrainian side
& 42 & 22& 23 &\textbf{95} &3 &1 &2 &1  & 10 & 1 \\
 4 & media, western media, journalism, news, reporting
& 25 & 10 & \textbf{27} &15&\textbf{27} &13 &14& 13  &23 & 15 \\ 
 5 & sanctions, individuals, entities, sanctions russian, legal entities
 & 21 & \textbf{37} & 3 &42 &9 &3 &7 & 6  &3 & 0 \\ 
  6 & join nato, join, membership, nato, ukraine membership
& \textbf{29} &14 &22 &26 &9 & 5 &6 & 3  & 2 & 0\\ 
 7 & kiev forces, kiev, units, people republic, kiev region & 12 & 7 &17 &7 &2  &1 &2 & 3 &\textbf{61} & 0\\
  8 & gas, russian gas, natural gas, energy, oil & 8 &17 &12 &7 &10 &  6&9&\textbf{23} & 7 & 1\\ 
 9 & demilitarization, denazification, demilitarization ukraine & 22 & 19 & 9& \textbf{29} &3 &4 &5 &{1}  &4 & 0 \\
 10 & evacuated, evacuation, city, leave, children evacuated
 & 12 &4 &20 &\textbf{21} &4 & 3&4 & 0 & 17 & 4\\
\end{tabular}
\caption{\label{tab:narratives} Top ten topics promoted by Russian state media websites relating to Ukraine along with the number of articles mentioning each topic from January 1 to April 5, 2022. The website with the most articles for each topic/narrative is bolded.} 
\vspace{-5pt}
\end{table*}

\vspace{-5pt}
\subsection{Growth and Spread of the Largest Narratives} 
\begin{figure}
  \centering
  \includegraphics[width=1.0\linewidth]{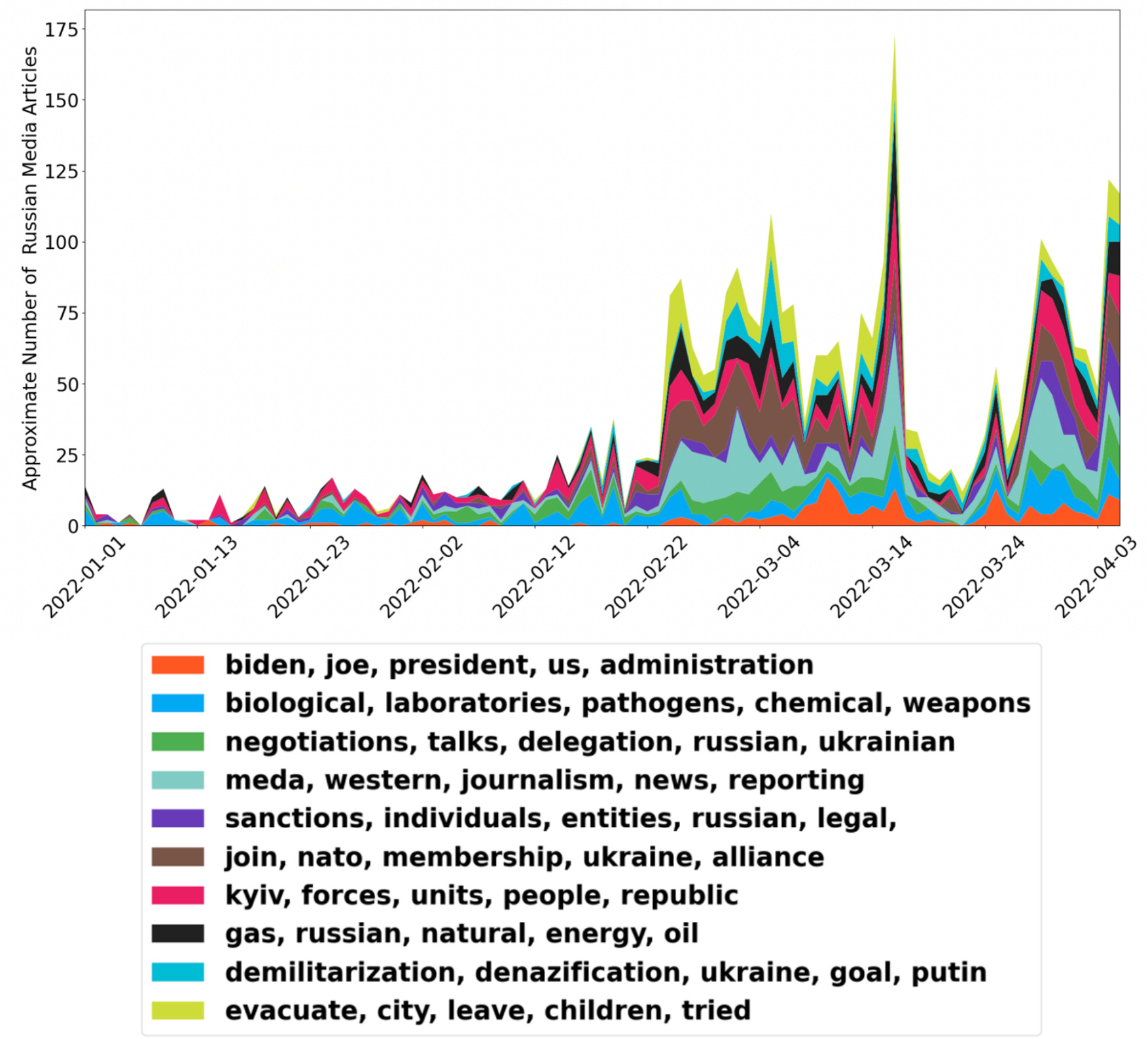}
\caption{Top ten topics (most frequently mentioned) on Russian state media websites relating to Ukraine from January 1 to April 5, 2022.
}
\label{fig:top-narrative-over-time}
\vspace{-10pt}
\end{figure}
In Table~\ref{tab:narratives}, we present the top 10 (most frequently written about) narrative clusters that Russian state-media websites promoted between January 1 and April 5, 2022. These narratives discuss a variety of topics relating to the Russo-Ukrainian War that range from biological weapons research and Russian gas to NATO expansionism. As can be seen in Figure~\ref{fig:top-narrative-over-time}, there are several peaks and troughs in the number of articles for some of the narratives, while other narratives/topics remain fairly constant. For example, articles about Joe Biden have remained fairly constant, while articles about biological weapons in Ukraine spiked after March 6. We explore some of these topics in some depth below:

\paragraph{Biological Weapons}
On March 6, 2022, the Russian news agency Tass reported that the US and Ukraine had attempted to eliminate samples of ``plague, anthrax, tularemia, cholera, and other deadly diseases'' from Ukraine prior to Russia's invasion on February 24.\footnote{\url{https://web.archive.org/web/20220412213821/https://tass.com/defense/1417951}} The accusation that the United States was helping fund biological weapons research in Ukraine was later echoed by other news reports across our dataset. Every website in our dataset wrote extensively about this topic in the forthcoming weeks. We see, in particular, 44~articles about the topic from Sputnik News, 37~from Tass, and 35~from Russia Today. We see this uptick most explicitly in Figure~\ref{fig:top-narrative-over-time}. We note that this narrative was thoroughly denied by the U.S. State Department~\cite{Price2022} and debunked by the New York Times~\cite{Qiu2022}. 

\paragraph{De-nazification of Ukraine}
In Russian President Vladimir Putin's announcement of the invasion of Ukraine, he stated that Russia's goal was to ``strive for the demilitarization and denazification of Ukraine''~\cite{Raghavan2022}. A major aspect of Russia's claim that Ukraine required ``denazification'' was that the Azov battalion volunteer force was a key part of Ukraine's military. The Azov Battalion is a para-military group launched by the ultranationalist group ''Patriot of Ukraine'' and the extremist group ``Social-National Assembly'' in 2014. Many, including the US government and the Ukrainian government, have attempted to moderate the group~\cite{Raghavan2022}. After being reorganized under the National Guard of Ukraine and after an effort in  2017, however, the Azov battalion has largely been considered to be depoliticized~\cite{Shekhovtsov2020}. Furthermore, despite the call from Vladimir Putin to ``denazify'' Ukraine, Ukraine's current president Volodymyr Zelensky is Jewish~\cite{Troianovski2022}. Further, while antisemitism remains a problem in Ukraine, according to polls conducted in 2016 by the Pew Research Center, Ukraine has some of the lowest rates of anti-Semitic attitudes in Eastern Europe~\cite{Masci2018}. We find that several websites in our dataset have written extensively about the Azov battalion with 29 Tass and 21 Russian Today articles mentioning the need to ``denazify'' Ukraine. 
\vspace{-5pt}
\subsection{War on Fakes Website}
Starting on March 4, 2022, waronfakes.com began utilizing ``fact-checking'' tactics to spread disinformation concerning the Russo-Ukrainian War (Figure~\ref{fig:waronfakes-examples}). As seen in Table~\ref{tab:narratives}, in particular, we find that five articles have mentions of biological weapons funded by the United States and four have mentions of ``misinformation in the reporting'' of Ukrainian evacuations of different cities. We note, however, that these topics are not the largest on the website.

One of the largest topics on waronfakes.com, mentioned by 58~articles, is the ``spread of misinformation by Ukrainians'' on Telegram and social media. As part of its ``fact-finding'' mission, the website in various articles cites how Ukrainians are spreading lies about the Russian atrocities occurring in Ukraine. For example, on March 28,  waronfakes.com ``fact-checked'' a rumor spreading on Telegram that the Russian military had destroyed a food depot.\footnote{\url{https://web.archive.org/web/20220409001851/https://waronfakes.com/mo-rf/fake-russian-troops-destroyed-a-food-storage-in-severodonetsk/}} Similarly, in response to information online about how the Russian military had burned a village down, waronfakes.com wrote an article denying it.\footnote{\url{https://web.archive.org/web/20220409002139/https://waronfakes.com/mo-rf/fake-russian-soldiers-drink-beer-after-burning-a-village/}}
Along these same lines, another one of the largest topics, containing 16 articles concerns correcting information about Russian destroyed buildings in Ukraine. A third topic with 15~articles discussed Western propaganda. This illustrates the extent to which waronfakes.com has targeted social media and Western media outlets in its ``fact-finding'' mission.

\begin{figure}
  \centering
  \includegraphics[width=1\linewidth]{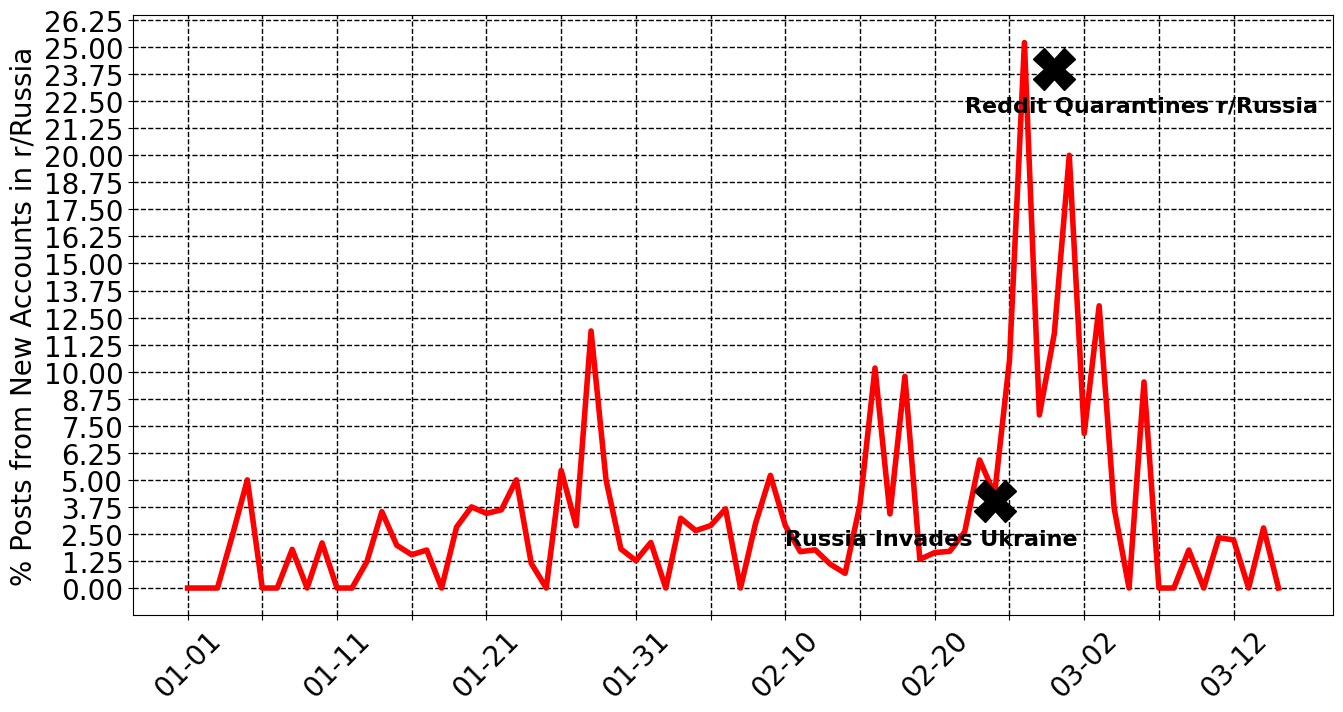}
  \caption{The percentage of \texttt{r/Russia} submissions that were posted by ``freshly created'' accounts (created within seven days of post).}
  \label{fig:percentage_of_new_users}
\vspace{-10pt}
\end{figure}

\vspace{-5pt}
\section{Russian State Media Narratives on Reddit}
In this section, we examine how the narratives spread by our set of  Russian state media websites interact with social media, specifically, the \texttt{r/Russia} subreddit. We examine this subreddit in particular because Reddit quarantined the subreddit due to the high degree of Russian disinformation present. Our approach aims to understand the extent to which \textit{specific} Russian state media narratives came to dominate conversations on Reddit and which narratives found traction on the social media platform.

To measure the spread of Russian state media narratives on \texttt{r/Russia}, we look at the set of English comments and submissions that were posted between January 1 and March 15, 2022. We detect the language of each comment using the Python \texttt{langdetect} library. We note here that because of the aggressive actions of Reddit when Russia invaded Ukraine, over 26,614 comments (24.9\%) were deleted or otherwise removed during this period. We thus are unable to comment and analyze this large subset of the comments on \texttt{r/Russia} and concentrate on the remaining comments.

\paragraph{\texttt{r/Russia} Subreddit Activity}
Before examining the narratives present in \texttt{r/Russia}, we note a surge in new accounts that were posted on the subreddit at the end of February. Upwards of one-fourth of the submissions on February 28 were posted by freshly created (created in the last 7 days) accounts (Figure~\ref{fig:percentage_of_new_users}). Examining the submissions made by users with freshly registered Reddit accounts, we find that many are pro-Russian and anti-Ukrainian. We list five of these submissions in Table~\ref{fig:submissions-comments}. This suggests the large degree to which users that previously did not post in the subreddit, as well as anti-Ukrainian narratives, came to be prevalent in the \texttt{r/Russia} subreddit before the community was quarantined. However, we further observe that following Reddit's quarantine of \texttt{r/Russia} on March 1, 2022, the number of daily comments and submissions in the subreddit decreased substantially; daily comments dropped from a seven-day average of 5,168 to 118.1, and daily submissions dropped from a seven-day average of 174.6 to 26. Reddit's quarantine effectively shut off conversation in the subreddit.

\paragraph{Mapping Reddit Comments} To understand if there is a correspondence between Reddit comments and our Russia state media topic clusters, we now map Reddit comments to the same 768-dimensional space as our news article sentences using the MPNet model from Section~\ref{sec:methodology}. We limit our study to comments with more than three words to ensure that each can be properly mapped and that each comment contains an interpretable topic; altogether 53,569~comments.

After mapping Reddit comments to the same dimensional space as our Russian article sentences, we utilize semantic search to find the cluster that is most similar to each Reddit comment. To do so, we average the set of sentence embeddings within each cluster to get an \textit{average cluster embedding}. Taking the cosine similarity of each Reddit comment to each \textit{average cluster embedding}, we find which cluster is most semantically similar to each Reddit comment. In order to properly map comments to different narrative clusters, we require that each comment is about the same topic as the cluster. Thus after finding the most similar cluster to each Reddit comment, only if the Reddit comment's similarity to that cluster is above a given threshold, do we assign that comment to the cluster. Given that our version of MPNet is fine-tuned for semantic search, by placing these comments in the same dimensional space as our set of news articles sentences, we can thus connect these comments to the set of articles/sentences that convey or talk about the same topic. 
\begin{table}
\centering
\small
\begin{tabular}{l}
\toprule
\textit{Dear kids, did you know that every time you say} \\\textit{``SLAVA UKRAINE!''}\\\textit{you actually say banderites salute?}\\ \textit{Oh, have you heard about those Ukrainian heroes?} \\ \textit{You should read more about it - it\'s very fascinating reading},\\\textit{ I guarantee it!} \\ \hline
\textit{According to Ukrainian propagandists},\\ \textit{Russian Military is already destroyed,}\\ \textit{and Ukrainian Military is ready to march to Moscow...}\\ \textit{just a bit later... a bit...'}\\\hline
\textit{Zelensky said that under martial law}\\ \textit{he will allow imprisoned people with combat experience}\\ \textit{to be released to help defend Ukraine}\\
\hline
\textit{Old Russian man shot for being possible Russian}\\
\textit{saboteur breaking curfew in Ukraine.}\\ \hline

\textit{EU bans Russian media for them telling Russian point of view. }\\ \textit{Also EU: ``Russia is an authoritarian country}\\\textit{where an alternative view isn\'t an option!.''}\\\textit{Wait, wait, I got it.}\\\textit{``Alternative view'' means EU sponsored? Right? Right? :)}\\ 
\hline
\end{tabular}
\caption{\label{fig:submissions-comments} \texttt{r/Russia} submissions made by ``freshly created'' accounts were pro-Russian and anti-Ukrainian.}
\vspace{-10pt}
\end{table}

\begin{figure}
\centering
  \begin{subfigure}{.35\textwidth}
  \includegraphics[width=1\linewidth]{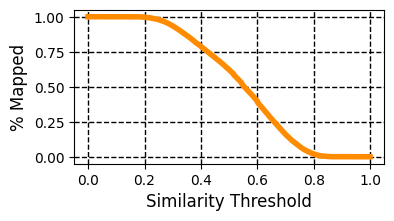}
  \end{subfigure}
\caption{The percentage of Reddit comments mapped to narrative clusters as a function of the similarity threshold.}
\label{fig:percentage-mapped}
\vspace{-10pt}
\end{figure}

\begin{table}
\centering
\fontsize{8.5pt}{9.0pt}
\selectfont
\setlength\tabcolsep{4pt}
\begin{tabular}{ar|ar|ar|ar}
 \multicolumn{2}{c|}{0.4 Sim.}  & \multicolumn{2}{c|}{0.5 Sim. }  & \multicolumn{2}{c|}{0.6 Sim. } &\multicolumn{2}{c}{0.7 Sim. } \\ \cline{1-8}\cline{1-8} 
   {\# Com.} & {Prec.} & {\# Com.}  & {Prec.}   & {\# Com.}  & {Prec.} & {\# Com.}  & {Prec.} \\ \hline
 88 & 100.0  & 87   &  100.0  & 78   &  100.0  & 45   &  100.0             \\ 
   54 & 24.0 &32 &40.6   & 13   &   69.2 & 5& 80.0           \\ 
  18 & 72.2   & 8   &  75.0 &   2   &  100.0 & 0 & ---               \\ 
 428 &95.1 & 373   & 96.5  & 259   & 96.9 &60 &100.0\\   
   34 & 91.2  & 34 & 91.2 & 32 & 96.9 &23 & 100.0     \\ 
 333 & 97.6 & 331   & 98.1 & 298 &100.0   & 159 &100.0         \\   
  10 & 100.0 & 9   & 100.0 & 9   & 100.0 &  6   & 100.0              \\   
 61 & 96.7  & 60   &  98.3   & 49   &  100.0 &  28   &  100.0                \\ 
  42 & 100.0  & 40   &  100.0  & 38   &  100.0     &26 &100.0            \\ 
 189 & 92.3  & 182 & 95.6 & 147   &99.3  & 67   &100.0              \\ 

 13 & 69.2 & 11   &81.8  & 8 & 87.5 & 4 & 100.0            \\ 
 17 & 94.1 & 12 & 100.0 & 8   &100.0 & 1   &100.0               \\ 
 223 & 91.0  & 202 & 93.1 & 131   &96.2  &35 &100.0             \\ 
  7 & 100.0 & 7 & 100.0 & 7   &100.0 &  7   &100.0              \\ 
 74 &98.6  & 72 & 100.0 & 64   &100.0   & 34   &100.0               \\ 
 11 & 81,8 & 7 & 100.0 & 5   &100.0     & 3   &100.0         \\ 
 58 & 93.1 & 57 & 94.7 & 52   &100.0   & 30   &100.0              \\ 
   3 & 66.6 & 3 & 66.6 & 2   &100.0  & 0   &---              \\ 
 16 & 75.0 & 15 & 80.0 & 3  &100.0 &    0   &---          \\   
  166 &97.6& 156 & 100.0 & 121  &100.0  & 43  &100.0          \\ \midrule 
 \multicolumn{1}{c}{Overall }   & 92.6& &95.6 & & 98.3& &\textbf{99.8} \\
\end{tabular}
\caption{\label{tab:precision} Evaluation of our methodology on 20 different topics (the top 10 topics and 10 random topics). } 
\vspace{-5pt}
\end{table}

\paragraph{Evaluation} We evaluate the precision of our approach in accurately mapping Reddit comments to state media narratives at different similarity thresholds. To perform our evaluation, we take the top 10~narrative clusters from Table~\ref{tab:narratives} as well as an additional 10 other topics and have an expert manually verify if the Reddit comments assigned to each cluster properly match each cluster's topic at the similarity thresholds of 0.4, 0.5, 0.6, and 0.7. We begin our threshold search at 0.40 as this indicates moderate similarity. As seen in Figure~\ref{fig:percentage-mapped}, this corresponds with 80\% of Reddit comments being mapped to a Russian state media topic cluster. We altogether examine 1,845 comments, 1,698 comments, 1,326 comments, and 563 comments across the 20 inspected topics at each threshold respectively. 

\begin{table}
\centering

\fontsize{9.0pt}{9.0pt}
\selectfont
\setlength\tabcolsep{4pt}
\begin{tabular}{lr}

Top Topics at 0.60 Threshold &  \# Com.  \\ \midrule
media,western media,journalism,news &  1038 \\ 
propaganda,russian propaganda,western propaganda,& 309\\ 
myth,countries demonstrated,revile,west follows, & 301\\ 
nato,new members,expansion,alliance,nato expansion,& 298\\ 
russia intend,attack anyone,anyone &281 \\ 
\end{tabular}
\caption{\label{tab:top-narratives-reddit} Top Five Topics/Narratives Connected to Russian state media narratives at a threshold of 0.60. }
\vspace{-10pt}
\end{table}

As seen in Table~\ref{tab:precision}, while the overall precision across the different topics at a threshold of 0.4 was 92.6\%, for certain topics, the approach's precision was faulty. At this threshold, the precision for Topic 2 was only 24.0\%, with the model assigning Reddit conversations about the Sputnik-IV Russian vaccine to this cluster about biological weapons. Similarly, for Topic 10 which concerns the Ukrainian evacuation of cities, most of the mislabeled sentences concerned other cities in the world that were in disrepair and were not worth visiting (according to the commenters). As the threshold increases, the precision of our approach increases at the expense of recall. Depending on our precision and recall needs, we find that we can thus adjust our filtering to achieve more accurate or more precise results.

In order to more conservatively label certain comments as belonging to the same topic as a given Russian state media narrative cluster, while maintaining high recall, we utilize a threshold of 0.6 for the rest of this work. At this threshold, we achieve an overall precision of 98.3\% as well as a precision of at least 69.2\% across every cluster inspected. This relatively high threshold further ensures that all comments are highly semantically similar to their assigned clusters. As previously noted in Section~\ref{sec:topics}, sentences with a similarity of nearly 0.60 largely contain similar semantic content; this threshold is further higher than the average inter-topic cluster similarity among the Russian news articles' sentences.

\begin{figure}
  \centering
  \includegraphics[width=1\linewidth]{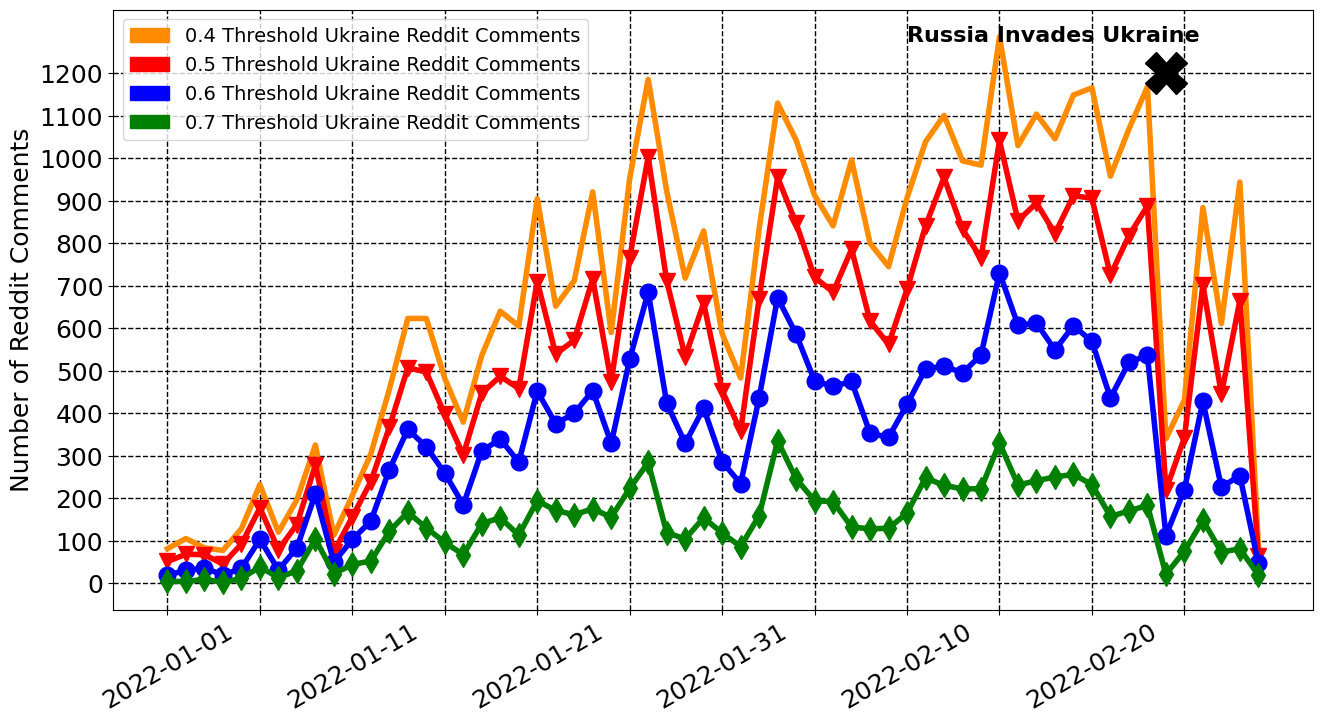}
\caption{Number of Russian-State Topic-Connected Reddit Comments over Time --- Throughout 2022, the number of comments posted  to \texttt{r/Russia} that were connected to Russian state media narrative increased steadily, only decreasing following Reddit's crackdown at the start of the Russo-Ukrainian War. 
}
\label{fig:ukraine-russian-narratives}
\vspace{-5pt}
\end{figure}

\paragraph{\texttt{r/Russia}'s Connection to Russian State Media}


 At the similarity threshold of 0.60, 21,250 (39.6\%) comments were mapped to the Russian 808 different state-media topics/narratives. Out of the 20,744 Reddit users who posted on \texttt{r/Russia} in our dataset, 8,184 (39.4 \%) users were responsible for these comments. Furthermore, just 819 (3.9\%) users were responsible for 50\% of these comments. We note, as we find 49 users with at least 100 comments connected to Russian state media, our approach can be utilized to discover and report users who are identified as pushers of specific Russian state-backed disinformation which we leave to future work.

 \begin{table}
\centering
\fontsize{9pt}{10pt}
\selectfont
\begin{tabular}{lr}
Domain & {0.6 \# Com. Origin Topics}\\ \midrule
rt.com & 5,113 \\
news-front.info &3,663 \\
strategic-culture.org& 3,474 \\
katehon.com & 3,501 \\
tass.com & 1,643 \\
geopolitica.ru & 1,264 \\
journal-neo.org&1,215  \\
southfront.org & 1,076 \\
sputniknews.com&  999 \\
waronfakes.com & 39 \\
\end{tabular}
\caption{\label{tab:comment-topics-originating} Number of comments assigned to topic/narrative clusters that each domain originated at the comment matching similarity threshold of 0.60. }  
\vspace{-10pt}
\end{table}

\begin{table*}
\centering
\fontsize{9.0pt}{9.0pt}
\selectfont
\setlength\tabcolsep{4pt}
\begin{tabular}{lrrlrr}
Subreddit  & \# Comm. & \% Russian   & Top Topic & \# Comm. about  & \# Comm. about  \\ 
           &          &    Narrative                   &           & Ukr. Bio. Lab &  Ukr. Nazis \\  \midrule
democrats & 25,314 & 8.82\% & biden, joe, joe biden, biden administration, president biden &16  &4  \\ 
geopolitics & 33,382 & \textbf{61.8\%} & geopolitial, clear warning, domain annexation, crimea & 9& 38 \\ 
socialism & 35,492 & 22.4\% & capitalist, bourgeois, marx, society, civil, proletarian & 0 &131 \\ 
republican &42,761 & 8.87\% & biden, joe, joe biden, biden administration, president biden & 2& 8\\ 
russia &53,569 & 39.6\% & biden, joe, joe biden, biden administration, president biden & 13& 147\\ 
libertarian & 321,439 & 9.66\%  &conservatism, democrats, says strategic, conservatism reflects &42 & 97 \\ 
ukpolitics & 517,487& 13.2\% & johnson, boris, boris johnson, scandals, farage& 34&115 \\ 
conservative &556,410& 10.4\% &biden, joe, joe biden, biden administration, president biden  & 212&79\\ 
canada & 862,485& 7.53\%&participating unauthorized, rallies, protest, government buildings & 22 &165\\ 
neoliberal &1,174,696 & 15.1\% & biden, joe, joe biden, biden administration, president biden & 556&153\\ 
politics & 1,747,381&8.86\% &biden, joe, joe biden, biden administration, president biden  &\textbf{1132}&\textbf{262}\\ 
\end{tabular}
\caption{\label{tab:all-top-narratives-reddit} Percentage of comments whose topics appeared in Ukraine-related Russian state news articles.}
\vspace{-5pt}
\end{table*}

As seen in Figure~\ref{fig:ukraine-russian-narratives}, the number of comments connected to Russian state media narratives increased steadily throughout 2022. Only following the invasion of Ukraine, when Reddit began to make a concerted effort to moderate the disinformation on \texttt{r/Russia} subreddit, did the number of comments connected to  Russian state-media narratives decrease. However, even given this massive drop following the Russian invasion of Ukraine, the number of comments connected to these narratives began to increase again before \texttt{r/Russia} was quarantined and the number of comments in the subreddit plummeted to near zero. 

In terms of the major narratives seen on the \texttt{r/Russia} subreddit, as seen in Table~\ref{tab:top-narratives-reddit}, some of the most popular Russian state-media sponsored topics on \texttt{r/Russia} were concerned with how  Western media and governments were demonizing the Russian government. Two topics in the top five narratives concern this idea. For example, one comment classified by our model called Western media ``\textit{Fake propaganda}''; another states ``\textit{Western aggression. Long live Russian people, Russian world. Western trolls eat dirt. Russia will win.}'' 

Looking at which website's originating topics (see Section~\ref{sec:topics}) had the largest impact on \texttt{r/Russia}, we observe in Table~\ref{tab:comment-topics-originating} that rt.com's originating topics had the most comments assigned to them (4,266), followed by news-front.info (2,753), strategic-culture.com (2,112), and katehon.com (2,072). We note that these are the same websites that we found in Section~\ref{sec:topics} with the largest reach in terms of originating content and in the external reposting of their original content. We thus observe that even though Reddit banned articles from Russian state media, specific topics pushed by different Russian media were still present on the \texttt{r/Russia} subreddit~\cite{Spangler2022}.


\paragraph{Political Subredddits' connection to Russian narratives}  We lastly analyze the degree to which the spread of narratives from Russian state media websites was localized to the \texttt{r/Russia} subreddit as opposed to the broader Reddit political ecosystem. In order to do so, we map comments posted between January 1, 2022, and March 15, 2022, from some of the largest political subreddits~\cite{rajadesingan2020quick} to our Russian media narrative clusters, again utilizing a 0.60 similarity threshold. Altogether we map an additional 5.37M Reddit comments across 10 different subreddits. We note this further illustrates the scalability of our approach to tracking different narratives across large social media ecosystems. To assess which cluster a comment corresponds to, we must  \textit{only} calculate its embedding's cosine similarity with each narrative cluster and then assess if the largest similarity is higher than our given threshold.  


As seen in Table~\ref{tab:all-top-narratives-reddit}, the degree to which Russian media narratives are associated with different subreddits varies widely. The top Russian-associated topic within each subreddit largely makes sense as well. Given that \texttt{r/geopolitics} largely discussed in detail the various aspects of the Russo-Ukrainian War, a geopolitical topic, we see that it has the highest percentage of its Reddit comments (61.8\%) that were associated with Russian media conversations about the war. \texttt{r/Russia} has the second-highest percentage with 39.6\% of its comments being associated with topics on Russian state media. In contrast, various other subreddits have much lower percentages of comments associated with narratives from Russian state media sites. \texttt{r/politics}, one of the largest subreddits discussing politics, has only 8.86\%, a far cry from 39.6\% in \texttt{r/Russia}. We leave identifying the set of all subreddits that have elevated levels of narratives associated with Russian state media websites to future work. 

Finally, to illustrate our methodology's ability to uncover and track disinformation, we track two disinformation narratives, US-funded Ukrainian bioweapons and elevated levels of Nazism within Ukraine, spread by Russian state media on each of these subreddits (detailed in Section~\ref{sec:topics}). As seen in Table~\ref{tab:all-top-narratives-reddit}, \texttt{r/politics} despite the lower percentage of comments associated with Russian disinformation, with the largest number of comments mapped, it has the largest absolute number of comments about both disinformation narratives. We find (we hypothesize due to the heavy moderation of \texttt{r/Russia}) that the \texttt{r/Russia} subreddit overall does not contain an outsized presence of these disinformation narratives.

\vspace{-10pt}
\section{Discussion and Conclusion}
On February 24, 2022, the Russian Federation invaded Ukraine with the stated goal to ``demilitarize and denazify'' the country. In this work, we utilize a fine-tuned version of the large language model MPNet to understand the narratives being spread by Russian state media websites surrounding the Russian invasion of Ukraine and their presence on the \texttt{r/Russia} subreddit. We discover that smaller websites like katehon.com, strategic-culture.org, news-front.info, and geopolitica.ru had an outsized effect in originating and propagating narratives within the Russian propaganda ecosystem, with other websites echoing the topics they introduce. These same websites' topics and narratives appeared the most frequently on the \texttt{r/Russia} subreddit, indicating their influence.

\vspace{-5pt}
\paragraph{Sentence Level Topic Analysis} In addition to performing an analysis of the key role of particular Russian propaganda websites, we show that a \textit{sentence-level} topic analysis approach can be used to identify and understand the presence of \textit{specific} Russian promoted narratives on  Reddit. A large insight of this approach relies on the idea of using \textit{sentence-level} topics. Ordinarily, topic modeling cannot be effectively computed on a sentence level due to word co-occurrence sparsity. To avoid this issue, we exploit large language models' ability to extract semantically rich embeddings~\cite{song2020mpnet}. With our approach, we successfully tracked Reddit comments within the \texttt{r/Russia} subreddit without relying on keywords or hyperlinks. We note that other ways of performing tracking such as utilizing keywords require a priori knowledge of specific disinformation narratives and can bias the results. This ability is part of what drives our approach's use of MPNet and BERTopic rather than LDA\@. Using the later approach to assign comments to topics would confine our work to matching keywords that occurred within clusters to keywords that occurred within comments. Approaches, built for document-level topic analysis like LDA further also assume that documents contain multiple topics; for social media posts like Reddit comments, this is largely not the case. Because the dictionaries between social media and news websites can radically differ despite discussing similar topics and due to LDA's breakdown on smaller texts, LDA is largely unsatisfactory for our purposes.

\vspace{-5pt}
\paragraph{Future Work} Our work can be extended to track and trace propaganda and media narratives across social media. By automatically identifying the set of promoted narratives present on a given website using clustering and then using these clusters to find similar narratives across sites like Reddit and Twitter, our approach can be utilized to understand the scale of influence of different ideas (disinformation or otherwise) across the Internet. In addition to identifying which state media narratives are most prominent on social media, we further note that our approach can also be utilized to identify which users on a platform are promoting specific narratives pushed by state media. By understanding which users are promoting given narratives and which \textit{specific} narratives are being promoted, subsequent action can be taken in accordance with each platform's values.
\vspace{-5pt}
\section*{Ethical Considerations}
We utilize only public data and follow ethical guidelines as outlined by others~\cite{hanley2021no}. We do not deanonymize users in our Reddit dataset, and our data collection does not breach the platform's terms of service. We recognize that Russo-Ukrainian War is an ongoing conflict and a humanitarian crisis. Sensitivity to the topic is paramount. We hope that our work provides objective insight into the information campaigns surrounding the war.

\vspace{-5pt}
\section*{Acknowledgements} 
This work was supported in part by the National Science Foundation under grant \#2030859 to the Computing Research Association for the CIFellows Project, a gift from Google, Inc., NSF Graduate Fellowship DGE-1656518, and a Stanford Graduate Fellowship.
\vspace{-5pt}
\bibliography{paper}

\end{document}